\documentclass[11pt]{article}
\usepackage{cite}
\usepackage{amsmath,amsfonts,amssymb,calrsfs}
\usepackage[small,bf,hang]{caption}

\def\hybrid{
        \topmargin -20pt
        \oddsidemargin 0pt
        \headheight 0pt \headsep 0pt
        \textwidth 6.25in 
        \textheight 9.5in 
        \marginparwidth .875in
        \parskip 5pt plus 1pt \jot = 1.5ex}

\hybrid

 \csname
@addtoreset\endcsname{equation}{section}

\def\moth{\mathsurround=0pt}
\newdimen\zo \zo=0pt

\def\tick{\leaders\hrule height 0.5ex depth 0pt \hskip 0.5pt}
\def\upboxfill{$\moth \setbox\zo\hbox{\tick}%
  \hskip 3pt\hbox to 0pt{$\tick$\hss}\hrulefill \hbox to 7.5pt{$\tick$\hss}$}

\def\dtick{\leaders\hrule height .34pt depth 0.5ex \hskip 0.5pt}
\def\downboxfill{$\moth \setbox\zo\hbox{\dtick}%
  \hskip 2pt\hbox to 0pt{$\dtick$\hss}\hrulefill \hbox to 2pt{$\dtick$\hss}$}

\def\bec{\begin{center}}
\def\ec{\end{center}}

\def\e{\epsilon}

\def\m{\mu}

\def\cE{{\cal E}}

\def\be{\begin{equation}}
\def\ee{\end{equation}}
\def\bea{\begin{eqnarray}}
\def\eea{\end{eqnarray}}
\def\ba{\begin{array}}
\def\ea{\end{array}}

\def\ft#1#2{{\textstyle{{\scriptstyle #1}
\over {\scriptstyle #2}}}}

\begin{document}

\begin{titlepage}
\begin{center}

\hfill UG-09-04 \\

\vskip 1.5 cm

{\LARGE \bf Can dual gravity be reconciled with $E_{11}$\;?
\\[0.2cm]}

\vskip 1.5cm

{\bf Eric A.~Bergshoeff, Mees de Roo and Olaf Hohm} \\

\vskip 25pt

{\em Centre for Theoretical Physics, University of Groningen, \\
Nijenborgh 4, 9747 AG Groningen, The Netherlands \vskip 5pt }

{email: {\tt E.A.Bergshoeff@rug.nl, M.de.Roo@rug.nl, O.Hohm@rug.nl}} \\

\vskip 0.8cm

\end{center}

\vskip 1cm

\begin{center} {\bf ABSTRACT}\\[3ex]

\begin{minipage}{13cm}
We extend a recently proposed formulation of dual gravity to the
case of eleven-dimensional supergravity. The supersymmetric action
corresponding to this alternative formulation is given, and it is
shown that it leads to a set of first-order duality relations from
which all second-order equations of motion follow as integrability
conditions. On top of the fields corresponding to the conjectured
$E_{11}$ symmetry the action features St\"uckelberg gauge fields
that facilitate the realization of arbitrary symmetries on the dual
graviton. However, there is no gauge-fixing that allows to eliminate
the St\"uckelberg fields. Therefore, $E_{11}$ by itself is not a
symmetry of eleven-dimensional supergravity but has to be extended
at least by the St\"uckelberg symmetries.

\end{minipage}

\vskip 3cm

\end{center}

\noindent

\vfill

March 2009

\end{titlepage}

\section{Introduction}\setcounter{equation}{0}
In recent years there has been increasing interest in a possible
relation between supergravity theories and Kac-Moody algebras. As
the most prominent example, it has been conjectured that the
infinite-dimensional Kac-Moody algebra $E_{11}$ is a symmetry of
11-dimensional supergravity and its lower-dimensional descendants
\cite{West:2001as,Schnakenburg:2001ya}. The main evidence for this
conjecture is given by the observation that the level decompositions
of $E_{11}$ with respect to its finite-dimensional subalgebras
reproduce the same p-form representations as expected for maximal
supergravity when formulated in a democratic way that introduces for
each p-form also its dual. For instance, the level decomposition of
$E_{11}$ with respect to $SL(11)$ reproduces a 3-form at level 1, in
agreement with the 3-form potential of 11-dimensional supergravity,
but also a 6-form at level 2. Though 11-dimensional supergravity
cannot be formulated entirely in terms of a dual 6-form, it is
possible to encode the dynamics in a democratic way in terms of a
first-order duality relation between the curvatures of the 3- and
6-form. Moreover, due to this appearance of extra gauge potentials
the gauge symmetry is enhanced in such a way that it permits a
global subgroup that, in turn, is in precise agreement with a
certain positive-level truncation of a non-linear realization of
$E_{11}$. To be specific, $E_{11}$ predicts the following global
symmetry on the 6-form,
 \bea\label{global}
  \delta A_{\mu_1\ldots \mu_6} \ = \ \Lambda_{\mu_1\ldots\mu_6}+
  20\Lambda_{[\mu_1\ldots\mu_3}A_{\mu_4\ldots\mu_6]}\;,
 \eea
while on the supergravity side there is a corresponding gauge
symmetry with parameters $\Lambda_{(2)}$ and $\Lambda_{(5)}$. The
latter reduces to the global symmetry (\ref{global}) upon specifying
the parameters to linear space-time dependence,
$\Lambda_{\mu_1\ldots\mu_5}(x)=\Lambda_{\rho\mu_1\ldots\mu_5}x^{\rho}$,
etc. Thus, there is a reformulation of the dynamics of the p-form
sector featuring the following two related properties:
\begin{itemize}
  \item[(i)]
   the second-order field equations result as integrability
   conditions from first-order duality relations between the physical fields
   and their duals predicted by $E_{11}$,
  \item[(ii)]
   the gauge symmetry is enhanced in agreement with
   $E_{11}$.
\end{itemize}

This correspondence between Kac-Moody algebras and (ungauged)
supergravities naturally extends to lower dimensions and fewer
numbers of supercharges. However, if one goes beyond low levels or
the pure p-form sector, the situation becomes more subtle. In the
$D=11$ decomposition, for instance, one finds a mixed-Young tableaux
representation at level 3, which is interpreted as the dual of the
graviton. Even though it is possible to formulate Einstein's
theory in the linearization about flat space entirely in terms of a dual
graviton \cite{Hull:2000zn,Hull:2001iu,West:2001as}, this turns out to be impossible for the non-linear theory
\cite{Bekaert:2002uh}. Also, a formulation in terms of a duality
relation is impossible even for linearized gravity once matter
fields are incorporated \cite{Bergshoeff:2008vc}. Finally, there is
no canonical way to associate to the rigid $E_{11}$ symmetry on the
dual graviton $C_{(8,1)}$, which is given by
 \begin{eqnarray}\label{dualsym}
  \delta C_{\mu_1\ldots\mu_8,\nu} &=&  \xi_{\mu_1\ldots\mu_8,\nu}
  +14 \Lambda_{\langle \mu_1\ldots\mu_6}A_{\mu_7\mu_8\nu\rangle}
  -14 A_{\langle\mu_1\ldots\mu_6}\Lambda_{\mu_7\mu_8\nu\rangle}\\
  \nonumber
  &&+\tfrac{7}{9} \Lambda_{\langle
  \mu_1\ldots\mu_3}A_{\mu_4\ldots\mu_6}A_{\mu_7\mu_8\nu\rangle}\;,
 \end{eqnarray}
a local gauge symmetry in supergravity. More precisely, due to the
non-trivial Young projection (here indicated by brackets
$\langle\;\rangle$) the global symmetry parameter cannot be
identified with the curl of a local parameter in a way that would
allow the definition of first-order curvatures.

Recently, a proposal has been made to overcome the problem, implied
by the no-go theorems of \cite{Bekaert:2002uh}, of finding an
equivalent reformulation of (super-)gravity that contains the dual
graviton and is valid at the non-linear level as well. Inspired by a
similar approach to gauged supergravity (see \cite{deWit:2008ta} and
references therein), an action has been given that contains the
original metric via a topological term and an additional shift gauge
field \cite{Boulanger:2008nd}. Moreover, in this formulation the
non-linear Einstein equations can be encoded in a set of
\textit{two} duality relations, thereby resolving the problems
mentioned above and preserving feature (i). This reformulation can
be investigated quite independently of $E_{11}$ and might be useful
for other applications as well.

The aim of the present letter is two-fold. First, in section 2, we
extend the proposed reformulation of \cite{Boulanger:2008nd} to the
special case of 11-dimensional supergravity. In particular, we will
show how all the symmetries of 11-dimensional supergravity, like
supersymmetry, are realized. This reformulation preserves feature
(i). Next, in section 3, we  address the question whether feature
(ii) is also preserved --- being a priori independent of (i) ---,
namely whether this reformulation of 11-dimensional supergravity
realizes the symmetries of $E_{11}$, in particular the one of the
dual graviton given in eq.~(\ref{dualsym}).

\section{An alternative formulation of 11-dimensional
supergravity}\label{sec:alternative}

In order to present the alternative formulation of 11-dimensional
supergravity, it is instructive to first show how the dual 6-form
potential is introduced. After that we will introduce the dual
graviton and, finally, we will discuss all the gauge symmetries of
the alternative formulation, including supersymmetry.

\subsection{Democratic formulation with 6-form
potential}\label{6formsec}

We start by giving a reformulation of
11-dimensional supergravity containing besides the standard fields,
i.e., the metric, the 3-form $A_{(3)}$ and the gravitino $\psi_{\mu}$, also the
dual 6-form potential $A_{(6)}$. It turns out that this is possible provided
one introduces in addition a 7-form gauge potential $Z_{(7)}$ that
gauges a shift symmetry on the 6-form. Thereby, the proper counting
of degrees of freedom will be maintained. The Lagrangian of
11-dimensional supergravity originally given in
\cite{Cremmer:1978km} reads\footnote{We follow the conventions of
\cite{Cremmer:1978km}, differing from those of
\cite{Boulanger:2008nd}. In particular, we choose the space-time
signature to be $(+-\ldots-)$, such that $\varepsilon^{01\ldots
11}=\varepsilon_{01\ldots 11}=+1$.}
 \bea\label{CJS}
  {\cal L} \ = \
  -e\;R-\tfrac{1}{12}\left(e\;F^{\mu\nu\rho\sigma}F_{\mu\nu\rho\sigma}
  -\tfrac{1}{216}\varepsilon^{\mu_1\ldots
  \mu_{11}}F_{\mu_1\ldots\mu_4}F_{\mu_5\ldots\mu_8}A_{\mu_9\mu_{10}\mu_{11}}\right)+{\cal
  L}_{\rm fermions}\;,
 \eea
where the field strength and gauge symmetry of the 3-form are given
by
 \bea
  F_{\mu\nu\rho\sigma} \ = \  4\partial_{[\mu}A_{\nu\rho\sigma]}\;,
  \qquad
  \delta A_{\mu\nu\rho} \ = \ 3\partial_{[\mu}\Lambda_{\nu\rho]}\;,
 \eea
and ${\cal L}_{\rm fermions}$ represents all terms containing the gravitino.

This action can be reformulated such that it contains a
kinetic term for the dual 6-form provided that  at the same time the 3-form
enters via an additional Chern-Simons-like topological coupling.
To be specific, we define the Lagrangian
 \bea\label{dualaction}
  {\cal L} \ = \ -e\;R
  +\tfrac{2}{7!}e\; F^{\mu_1\ldots\mu_7}F_{\mu_1\ldots\mu_7}
  +{\cal L}_{\rm top}+{\cal
  L}_{\rm fermions}\,,
 \eea
where the topological terms are given by
 \bea\label{top}
  {\cal L}_{\rm top} \ = \
  -\tfrac{2}{3\cdot 7!}\varepsilon^{\mu_1\ldots\mu_{11}}
  Z_{\mu_1\ldots\mu_7}\partial_{\mu_8}A_{\mu_9\mu_{10}\mu_{11}}
  +\tfrac{1}{12\cdot 216}\varepsilon^{\mu_1\ldots
  \mu_{11}}F_{\mu_1\ldots\mu_4}F_{\mu_5\ldots\mu_8}A_{\mu_9\mu_{10}\mu_{11}}\;.
 \eea
Here, we have
defined  the field strength of the 6-form as follows
 \bea
  F_{\m_1\ldots\m_7} \ = \ 7\,\partial_{\,[\,\m_1}A_{\m_2\ldots\m_7]}
   + Z_{\m_1\ldots\m_7}\;,
 \eea
such that it is invariant under the local St\"uckelberg shift
symmetry
 \bea\label{localshift}
  \delta A_{\mu_1\ldots\mu_6} \ = \ -\Sigma_{\mu_1\ldots\mu_6}\;,
  \qquad
  \delta Z_{\mu_1\ldots\mu_7} \ = \
  7\partial_{[\mu_1}\Sigma_{\mu_2\ldots\mu_7]}\;.
 \eea
The newly introduced 7-form $Z_{(7)}$ acts as the shift gauge field.

The theory defined by (\ref{dualaction}) is on-shell equivalent to
the original action (\ref{CJS}). The easiest way to see this is to
use the local St\"uckelberg symmetry (\ref{localshift}) to gauge-fix
$A_{(6)}$ to zero and then to integrate out $Z$. By virtue of the
topological term (\ref{top}) this results in the proper kinetic term
for the original 3-form in (\ref{CJS}). Note that this equivalence
is not affected by the precise form of the fermionic couplings and
therefore supersymmetry extends to (\ref{dualaction}), whose
realization we will discuss in sec.~\ref{gauges} in more detail.

At this stage one may wonder whether it is not artificial to
introduce the 6-form together with a local shift symmetry such that
it can be gauged away completely. However, apart from the fact that
it needs to be possible to eliminate $A_{(6)}$ in order to guarantee
the equivalence to the original formulation without a 6-form, it is
precisely this framework that allows us to analyze the most general gauge
symmetries on the 6-form. Since a shift symmetry
is the largest possible gauge symmetry, any other gauge invariance
consistent with the dynamics of 11-dimensional supergravity has to
result from (\ref{localshift}) by a gauge-fixing. In particular, we will show
how the $E_{11}$ structure indeed arises through a gauge-fixing of
(\ref{localshift}) that is different from gauging $A_{(6)}$ away.

To start with, we note that the reformulation \eqref{dualaction}
preserves feature (i), i.e.~the field equations for $Z_{(7)}$ and
$A_{(3)}$ are given by the following two first-order `duality
relations'
 \be
   F_{\m_1\ldots \m_7} = \tfrac{1}{4!}\,\varepsilon_{\m_1\ldots\m_7\lambda_1\ldots\lambda_4}
   F^{\lambda_1\ldots \lambda_4} \,,
\label{dualrel} \ee and
 \bea\label{dualrel2}
  \partial_{[\mu_1}Z_{\mu_2\ldots\mu_8]} \ = \ \tfrac{35}{4}
  F_{[\mu_1\ldots\mu_4}F_{\mu_5\ldots\mu_8]}+{\rm fermions}\;.
 \eea
The second-order field equation of $A_{(6)}$ can be obtained from
(\ref{dualrel}) by acting with a derivative. Similarly, the
second-order equations for $A_{(3)}$ corresponding to the original
action (\ref{CJS}) can be obtained by first taking the exterior
derivative of (\ref{dualrel}) and using the Bianchi identity
 \bea\label{bianchiZ}
  \partial_{[\mu_1}F_{\mu_2\ldots\mu_8]} \ = \
  \partial_{[\mu_1}Z_{\mu_2\ldots\mu_8]}\;,
 \eea
and next applying the second duality relation (\ref{dualrel2}). Thus, the
second-order field equations for the p-form sector of 11-dimensional
supergravity can be obtained as the integrability conditions of a
set of two first-order `duality' relations. As we will see below,
this is a rather general feature that also holds for the gravity
sector.

We now show how in the present case the relation to $E_{11}$ emerges
by a particular gauge-fixing. We first write the right-hand side of
(\ref{dualrel2}) as the exterior derivative of $A_{(3)}\wedge
F_{(4)}$, where we momentarily ignore fermionic terms. Consequently,
(\ref{dualrel2}) can be locally solved by virtue of the Poincar\'e
lemma, implying
 \bea\label{Zsol}
  Z_{\mu_1\ldots\mu_7} \ = \ 35\,
  A_{[\mu_1\ldots\mu_3}F_{\mu_4\ldots\mu_7]}+\partial_{[\mu_1}\Xi_{\mu_2\ldots\mu_7]}\;.
 \eea
Here, a new 6-form $\Xi$ arises, to which we have to assign the
following non-trivial gauge transformation in order for $Z$ to
transform as required by (\ref{localshift}):
 \bea
  \delta \Xi_{\mu_1\ldots\mu_6} \ = \ 7\,\Sigma_{\mu_1\ldots\mu_6}
  +42 \,\partial_{[\mu_1}\Lambda_{\mu_2\ldots\mu_6]}
  -105\,\Lambda_{[\mu_1\mu_2}F_{\mu_3\ldots\mu_6]}\;.
 \eea
Here $\Lambda_{(5)}$ is a new gauge parameter leaving (\ref{Zsol})
invariant. Since $\Xi$ transforms by a shift under the St\"uckelberg
symmetry, we can gauge-fix this symmetry by setting $\Xi=0$. This in
turn requires compensating gauge transformations with parameter
 \bea
  \Sigma_{\mu_1\ldots\mu_6} \ = \
  -6\,\partial_{[\mu_1}\Lambda_{\mu_2\ldots\mu_6]}+
  15\,
  \Lambda_{[\mu_1\mu_2}F_{\mu_3\ldots\mu_6]}\;,
 \eea
leaving the following gauge symmetry on $A_{(6)}$ as the remnant of
the shift symmetry (\ref{localshift})
 \bea\label{6formgauge}
  \delta A_{\mu_1\ldots\mu_6} \ = \ 6\,\partial_{[\mu_1}\Lambda_{\mu_2\ldots\mu_6]}
  -15\,
  \Lambda_{[\mu_1\mu_2}F_{\mu_3\ldots\mu_6]}\;.
 \eea
Upon redefining the parameter $\Lambda_{(5)}$, this transformation
rule can be brought into a form in which the symmetry parameters
appear only under a derivative,
 \bea
  \delta A_{\mu_1\ldots\mu_6} \ = \
  6\left(\partial_{[\mu_1}\Lambda_{\mu_2\ldots\mu_6]}+10\partial_{[\mu_1}\Lambda_{\mu_2\mu_3}
  A_{\mu_4\ldots\mu_6]}\right)\;,
 \eea
which reproduces the $E_{11}$ structure (\ref{global}) by choosing
 \bea
  \Lambda_{\mu\nu} \ = \ \ft13 \Lambda_{\mu\nu\rho}x^{\rho}\;,
  \qquad
  \Lambda_{\mu_1\ldots\mu_5} \ = \
  \tfrac{1}{6}\Lambda_{\rho\mu_1\ldots\mu_5}x^{\rho}\;.
 \eea
Moreover, after insertion of (\ref{Zsol}), the 7-form field strength
reduces to
 \bea
  F_{\mu_1\ldots\mu_7} \ = \ 7\,\partial_{\,[\,\m_1}A_{\m_2\ldots\m_7]}
   +35\, A_{[\mu_1\ldots\mu_3}F_{\mu_4\ldots\mu_7]}\;,
 \eea
which is invariant under (\ref{6formgauge}) and corresponds to a
Maurer-Cartan form of the non-linear realization of
$E_{11}$.\footnote{There exists a slightly different way of
presenting the action, in which the $E_{11}$ structure appears
already at the level of the action before gauge-fixing. For this one
needs to redefine $Z$ by a term $A_{(3)}dA_{(3)}$ in such a way that
the Chern-Simons-like structure appears inside the 7-form field
strength and not as the usual topological term in the action. The
latter is then generated upon integrating out $Z$.} After this
gauge-fixing the first duality relation (\ref{dualrel}) already
encodes the full dynamics since the right-hand side of the Bianchi
identity (\ref{bianchiZ}) gets replaced by the right-hand side of
(\ref{dualrel2}). Upon including the fermions in this analysis it is
still possible to solve for $Z$ since $A_{(3)}$ couples only via the
gauge-invariant field-strength $F_{(4)}$ to the fermions. As a
consequence, the field strengths in (\ref{dualrel}) will be replaced
by  supercovariant curvatures $\hat F$, in agreement with
supersymmetry. It is this formulation in terms of a single duality
relation which is usually presented in order to relate the p-form
gauge symmetries of supergravity to $E_{11}$. However, this
procedure leading to a single duality relation cannot be implemented
at the level of the action, since derivatives were involved when
solving some of the field equations. Therefore, the field $Z$ is
indispensable in order to obtain the first-order duality equations
from an action. We will now discuss a similar reformulation with
St\"uckelberg gauge fields that involves dual gravity.

\subsection{Democratic formulation with dual graviton}

In order to introduce the dual graviton, it is convenient to use
that the Einstein-Hilbert action in $D=11$ can be written, up to
total derivatives, in the following first-order form
\cite{West:2001as,West:2002jj}
 \bea\label{firstorder}
  {\cal L}_{\rm EH} \ = \
  \frac{1}{2}e\left(Y^{ab|c}\Omega_{abc}-\frac{1}{2}Y_{ab|c}Y^{ac|b}
  +\frac{1}{18}Y_{ab|}{}^{b}Y^{ac|}{}_{c}\right)\;,
 \eea
where
 \bea
  \Omega_{ab}{}^{c} \ = \
  e_{a}{}^{\mu}e_{b}{}^{\nu}\left(\partial_{\mu}e_{\nu}{}^{c}-\partial_{\nu}e_{\mu}{}^{c}\right)
 \eea
are the coefficients of anholonomy, and $Y_{ab|c}=-Y_{ba|c}$ is an
auxiliary field with no further symmetry properties, i.e.,
transforming in a reducible representation. Integrating out $Y$ one
recovers the Einstein-Hilbert action written through $\Omega^2$
terms. Below we will use that any symmetry of the original
Einstein-Hilbert action can be extended to an (off-shell) invariance
of the first-order action (\ref{firstorder}) by assigning a suitable
transformation rule to $Y_{ab|c}$. More precisely, an arbitrary
symmetry of the Einstein-Hilbert action with transformations $\delta
e_{\mu}{}^{a}$ is elevated to a symmetry of (\ref{firstorder}) with
$\delta Y_{ab|c}$ given by \bea\label{offshellY}
  \delta Y_{ab|c} \ = \ \delta
  \Omega_{abc}-2\delta\Omega_{c[ab]}+4\eta_{c[a}\delta\Omega_{b]d}{}^{d}\;.
 \eea

We next define the Hodge dual via
 \bea\label{dualY}
  Y^{ab|c} \ = \ \tfrac{1}{9!}\epsilon^{abc_1\cdots c_{9}}
  Y_{c_1\cdots c_{9}|}{}^{c}\;,
 \eea
which yields a $(9,1)$ tensor and which will below play the same
role as the 7-form shift gauge field $Z$ in the previous subsection.
We now consider the Lagrangian
 \bea\label{finalaction}
  {\cal L} \ = \ {\cal L}_{\rm C}(e,G)
  +\tfrac{2}{7!}e\, F^{\mu_1\ldots\mu_7}F_{\mu_1\ldots\mu_7}
  +\widehat{{\cal L}}_{\rm top}+{\cal
  L}_{\rm fermions}\;,
 \eea
where  ${\cal L}_{\rm C}(e,G)$ is in form equal to the so-called
Curtright Lagrangian for the dual graviton $C_{\mu_1\cdots
\mu_8}{}^a$,
 \begin{eqnarray}
   {\cal L}_{\rm C}(e,G) &=& -\tfrac{1}{9!}\big(\tfrac{2}{9}\,e\,
   G^{\mu_1\ldots\mu_{9} a}G_{\mu_1\ldots\mu_{9} a}
   -\tfrac{9}{4}\;e\, e_{\nu}{}^{a}\,e_{b}{}^{\rho}\,
   G^{\mu_1\ldots\mu_{8}\nu}{}_{a}\,
   G_{\mu_1\ldots\mu_{8}\rho}{}^{b}
    \nonumber \\
   && +\tfrac{1}{4}\,e\,e_{\nu}{}^{b}\,e_{a}{}^{\rho}\,
   G^{\mu_1\ldots\mu_{8}\nu a}\,G_{\mu_1\ldots\mu_{8}\rho b}\big)
   \;.
 \end{eqnarray}
(We will discuss in sec.~\ref{E11sym} why it is justified to call
$C$ the `dual graviton'.) To be precise, while the conventional
Curtright action \cite{Curtright:1980yk,Aulakh:1986cb} for the dual
graviton is formulated on flat space, here we keep the dynamical
metric in order to maintain full diffeomorphism invariance. Another
difference with the Curtright action is that the field strength
entering here is the shift-invariant combination
 \bea
  G_{\mu_1\ldots\mu_{9}}{}^{a} \ = \
  9\;\partial_{[\mu_1}C_{\mu_2\ldots\mu_{9}]}{}^{a}+Y_{\mu_1\ldots\mu_9}{}^{a}\;,
 \eea
admitting the symmetry
 \bea\label{shiftsym}
  \delta Y_{\mu_1\cdots\mu_{9}}{}^{a} \ = \
  9\;\partial_{[\mu_1}\Sigma_{\mu_2\cdots\mu_{9}]}{}^{a}\;,
  \qquad
  \delta C_{\mu_1\cdots\mu_{8}}{}^{a} \ = \
  -\Sigma_{\mu_1\cdots\mu_{8}}{}^{a}\;.
 \eea
Moreover, the topological couplings in (\ref{top}) have been
extended by a term involving the original vielbein $e_{\mu}{}^{a}$,
 \bea
  \widehat{{\cal L}}_{\rm top} \ = \ {\cal L}_{\rm top}
  +\tfrac{1}{9!}
  \varepsilon^{\mu_1\ldots\mu_9\nu\rho}\;Y_{\mu_1\ldots\mu_9\;a}\;\partial_{\nu}e_{\rho}{}^{a}\;.
 \eea
We are working here in a frame-like formulation (with flat indices
$a,b,\ldots$) for which the dual graviton lives in a reducible
representation. The antisymmetric indices are chosen to be curved,
while the extra index is flat. This assignment is natural in that it
keeps the diffeomorphism symmetry manifest, leaving the local
Lorentz group as the only non-manifest symmetry
\cite{Boulanger:2008nd}.

It can now be shown in complete analogy to the discussion in the
previous subsection that the action (\ref{finalaction}) containing
all fields required by $E_{11}$ is on-shell equivalent to
11-dimensional supergravity.  To show this equivalence one may
gauge-fix the dual graviton to zero, after which the terms
containing $Y$ are given by (\ref{firstorder}) when rewritten
according to (\ref{dualY}). Thus, it is equivalent to the
Einstein-Hilbert action. The analysis of the foregoing section
concerning the 3-form/6-form sector is unaffected, and so
(\ref{finalaction}) is equivalent to 11-dimensional supergravity.

Let us now inspect the equations of motion. Varying with respect to
$Y$ we obtain a duality relation between the original vielbein and
the dual graviton,
 \begin{eqnarray}\label{dualitygrav}
  e^{-1}\varepsilon^{\mu_1\ldots\mu_{9}\nu\rho}\Omega_{\nu\rho}{}^{a}
  &=& \tfrac{8}{9}\,G^{\mu_1\ldots\mu_{9}\;a}
  -9\,e_{\rho b}e^{a[\mu_1}G^{\mu_2\ldots\mu_{9}]\rho\;b}
  +e_{\rho}{}^{a}e_{b}{}^{[\mu_1}G^{\mu_2\ldots\mu_{9}]\rho\;b}\;,
 \end{eqnarray}
while variation with respect to the vielbein yields
 \bea\label{2ndduality}
  -\tfrac{1}{9!}e^{-1}\varepsilon^{\mu\mu_1\ldots\mu_{10}}\partial_{\mu_1}
  Y_{\mu_2\ldots\mu_{10}\;a}
  \ = \  e^{-1}{\cal T}^\mu{}_a \ \equiv\  \,e^{-1}\frac{\delta{\cal L}_{\rm C}(e,G)}
  {\delta e_{\mu}{}^{a}}+e^{-1}T^{\mu}{}_{a}\;,
 \eea
where $T^{\mu}{}_{a}$ denotes the energy-momentum tensor of
$A_{(6)}$. As in the 3-form/6-form example, the second-order field
equations for the dual graviton $C$ can be obtained from
(\ref{dualitygrav}) by taking the exterior derivative. Therefore,
the full set of field equations including the non-linear Einstein
equations is encoded in the two first-order `duality' relations
(\ref{dualitygrav}) and (\ref{2ndduality}), preserving feature (i),
cf.~the introduction, at the non-linear level. In particular, this
circumvents the problem that it is not possible to `pull out' a
derivative of the energy-momentum tensor and that it is, therefore,
not possible to encode matter couplings in a single duality relation
\cite{Bergshoeff:2008vc}. The way out is to introduce a new gauge
field $Y$ together with a second duality relation. We stress that
the need of having \textit{two} duality relations is by no means a
peculiarity of gravity. For instance, for obtaining the scalar field
equations in gauged supergravity from first-order duality relations
one is confronted with the analogous problem that it is not possible
to `pull out' a derivative from the source term induced by the
scalar potential. The resolution is, again, to introduce a second
duality relation involving a higher-rank p-form
\cite{Bergshoeff:2009ph}, in accordance with the so-called  tensor
hierarchy \cite{deWit:2008ta}.

\subsection{Gauge symmetries}\label{gauges}

In this subsection we are going to show how in the proposed
reformulation of $D=11$ supergravity  the
 symmetries of the original 11-dimensional supergravity theory, as for instance supersymmetry, are realized.
Due to the on-shell equivalence of the two formulations, the
existence of these symmetries in a local form is guaranteed. It is,
however, instructive to determine them explicitly.\footnote{In the
context of $N=1$ supergravity in $D=4$ the supersymmetry of this
reformulation has been analyzed in \cite{Nurmagambetov:2008yg},
however, with inconclusive results.}

We focus first  on the p-form sector only, i.e., we assume that in
the gravitational sector only the ordinary metric enters, via the
standard Einstein-Hilbert action. The following discussion shows how
any symmetry of 11-dimensional supergravity can be elevated to a
symmetry of the reformulation. We first note that the variation of
the original kinetic term for the 3-form reads
 \begin{eqnarray}\label{totvar}
  \delta {\cal L}_{\rm 3-form} &=&
  \tfrac{2}{3} \,\delta A_{\mu_1\ldots\mu_3}
  \partial_{\rho}F^{\rho\mu_1\ldots\mu_3} \\ \nonumber
  &&-\tfrac{1}{12}\,\delta g^{\mu\nu}\left(4\sqrt{g}F_{\mu}{}^{\rho_1\ldots\rho_3}F_{\nu\rho_1\ldots\rho_3}
  -\tfrac{1}{2}\sqrt{g}g_{\mu\nu}F^{\mu_1\ldots\mu_4}F_{\mu_1\ldots\mu_4}\right)\;.
 \end{eqnarray}
In  the case of supersymmetry, the variations of the elf-bein and the 3-form are given by
 \bea
  \delta e_\m{}^a &=& \tfrac{1}{2}\,\bar\e\,\gamma^a\psi_\m\;,
  \qquad
  \delta A_{\mu_1\ldots\mu_3} \ = \
  \tfrac{3}{2}\;\bar{\epsilon}\gamma_{[\mu_1\mu_2}\psi_{\mu_3]}\;.
 \eea
One may verify that the variation of the kinetic term for the 6-form
and the additional topological term containing $Z$ precisely
reproduces (\ref{totvar}) provided we assign the following
transformation rules to the new fields
 \begin{eqnarray}\label{newvar}
  \delta Z_{\mu_1\ldots\mu_7} &=& \tfrac{1}{4!}\varepsilon_{\mu_1\ldots\mu_7\nu_1\ldots\nu_4}
  \delta F^{\nu_1\ldots\nu_4}
  +\tfrac{1}{4}\;g_{\mu\nu}\delta g^{\mu\nu}\cE^{+}_{\mu_1\ldots\mu_7}
  +\tfrac{7}{2}\;\delta
  g^{}_{\nu[\mu_1}\cE_{+\mu_2\ldots\mu_7]}^{\nu}\;, \\
  \delta A_{\mu_1\ldots\mu_6} &=&  0\;,
 \end{eqnarray}
while the old ones remain unmodified. Here we have defined
 \bea\label{chi}
  {\cal E}_{\pm}{}^{\mu_1\ldots\mu_7} \ = \ F^{\mu_1\ldots\mu_7}\pm
  \tfrac{1}{4!}\varepsilon^{\mu_1\ldots\mu_7\nu_1\ldots\nu_4}F_{\nu_1\ldots\nu_4}\;,
 \eea
of which the $\cE_{-}$ combination represents the duality relation,
i.e., vanishes on-shell, $\cE_{-}=0$, while the other combination is
on-shell given by the field strength, $\cE_{+}=2F_{(7)}$. It is
amusing to note that in this sense precisely the `opposite' of the
duality relation enters in (\ref{newvar}). In total we have shown
that (\ref{dualaction}) is invariant under the combined
transformation of the old fields and (\ref{newvar}). The fact that
the symmetry transformation on $A_{(6)}$ can be taken to vanish is
due to the St\"uckelberg invariance. Once the latter is gauge-fixed
as in sec.~\ref{6formsec}, compensating gauge transformations are
required, which in turn give rise to non-trivial transformations of
$A_{(6)}$. In the case of supersymmetry, this procedure leads to the
following supersymmetry rule:
 \bea
  \delta A_{\mu_1\ldots\mu_6} \ = \
  3\;\bar{\epsilon}\gamma_{[\mu_1\ldots\mu_5}\psi_{\mu_6]}\;,
 \eea
where we ignore higher-order terms. Here we used the field
equations, i.e., this supersymmetry rule holds on-shell.

Similar conclusions apply to the gravitational sector once the dual
graviton is introduced in the Lagrangian (\ref{finalaction}). In
fact, in (\ref{offshellY}) we have already given the off-shell
symmetry relevant for this reformulation, which determines the
symmetry rules for the shift gauge field $Y$. Due to the
St\"uckelberg invariance, the dual graviton can be taken to be
invariant, and it will only start transforming after a gauge-fixing.

\section{Can the gauge symmetries be reconciled with
$E_{11}$?}\label{E11sym}

So far we have shown that it is possible to
reformulate the action of 11-dimensional supergravity in such a way
that it contains the fields required by $E_{11}$ at low levels
together with two extra St\"uckelberg gauge fields. Moreover, the
field equations can be encoded in a set of first-order `duality'
relations, in agreement with feature (i) mentioned in the introduction.
In this section, we are going to investigate to what extent this
realizes also the $E_{11}$ symmetry, i.e., whether at the same time the reformulation
preserves feature (ii).

We first discuss in which sense it is justified to call $C$ the
`dual graviton'. Even though $C$ can be gauged away at any step,
there is a limit in which there is a different gauge-fixing, giving
rise to a propagating dual graviton. More precisely, in the
linearization about flat space gravity decouples from matter and
thus the second duality relation (\ref{2ndduality}) reduces to
$dY^{a}=0$. Since so far $Y^{a}$ does not transform in a specific
Young tableaux representation, the index $a$ can be treated as a
redundant index. Thus, applying the Poincar\'e lemma, the shift
gauge field is determined to be pure gauge,
 \bea\label{simpdual}
  \delta Y_{\mu_1\ldots\mu_9}{}^{a} \ = \ 9\;
  \partial_{[\mu_1}\Xi_{\mu_2\ldots\mu_9]}{}^{a}\;, \qquad
  \delta \Xi_{\mu_1\ldots\mu_8}{}^{a} \ = \
  \partial_{[\mu_1}\gamma_{\mu_2\ldots\mu_8]}{}^{a}+\Sigma_{\mu_1\ldots\mu_8}{}^{a}\;,
 \eea
and can therefore be gauge-fixed to zero. By (\ref{simpdual}) this
requires compensating gauge transformations on the dual graviton,
giving rise to
 \bea\label{dualdiff}
  \delta_{\gamma}C_{\mu_1\ldots\mu_8}{}^{a} \ = \
  \partial_{[\mu_1}\gamma_{\mu_2\ldots\mu_8]}{}^{a}\;.
 \eea
These are the `dual diffeomorphisms', while the action reduces to
the Curtright action for the dual graviton, being invariant under
(\ref{dualdiff}). To be more precise, the theory still has a
local Lorentz symmetry which makes it possible to gauge away the totally
antisymmetric part of $C$, leaving the dual graviton in the (8,1)
Young tableaux representation \cite{Boulanger:2008nd}.
Correspondingly, (\ref{dualdiff}) reduces to a (non-manifest) gauge
symmetry with two parameters in irreducible
representations,\footnote{In (\ref{irresym}) we use that on flat
space there is no distinction between flat and curved indices. }
 \bea\label{irresym}
  \delta C_{\mu_1\ldots\mu_8,\nu} \ = \
  \partial_{[\mu_1}\alpha_{\mu_2\ldots\mu_8],\nu}+\partial_{\langle
  \mu_1}\beta_{\mu_2\ldots\mu_8\nu\rangle}\;,
 \eea
where $\alpha$ transforms in the $(7,1)$ tableaux and $\beta$ is
fully antisymmetric. This gauge symmetry of the dual graviton can be
associated to the global shift transformation predicted by
non-linear realizations of Kac-Moody algebras. More precisely,
choosing
 \bea
  \alpha_{\mu_1\ldots\mu_7,\nu} \ = \
  \xi_{\mu_1\ldots\mu_7\rho,\nu}x^{\rho} \qquad \text{or } \qquad
  \beta_{\mu_1\ldots\mu_8} \ = \
  \xi_{\mu_1\ldots\mu_8,\nu}x^{\nu}\;,
 \eea
one recovers the global symmetry transformation encoded in the first
term of (\ref{dualsym}), which is precisely the symmetry predicted
for pure 11-dimensional gravity based on the Kac-Moody algebra $A_{8}^{+++}$.

We now turn to the question whether also the $E_{11}$ structure
going beyond the dual diffeomorphisms, i.e.~the remaining terms in
the transformation rule \eqref{dualsym}, can be obtained in this
way. First we observe that due to the St\"uckelberg symmetry
(\ref{shiftsym}) \textit{any} symmetry can be realized on the dual
graviton by simply choosing the shift parameter $\Sigma$ in the
required way, as for instance suggested by the $E_{11}$ structure
(\ref{dualsym}). However, this trivial way of realizing $E_{11}$  is
clearly unsatisfactory. As in our previous discussions, what we
really have to ask for is a gauge--fixing that allows to eliminate
the St\"uckelberg gauge field $Y$ in such a way that the residual
gauge symmetry on the dual graviton is given by (\ref{dualsym}). For
this to be the case, it has to be possible to solve the second
duality relation (\ref{2ndduality}) for $Y$ up to pure gauge degrees
of freedom. As the two terms appearing on the right-hand side of
that duality relation can be interpreted as the energy-momentum
tensor of the dual graviton and the 6-form, respectively, this
problem is similar to the one encountered in
\cite{Bergshoeff:2008vc} of pulling out a derivative of the
energy-momentum tensor, which turned out to be impossible. Even
though here the situation is slightly different in that we are not
dealing with the ordinary energy-momentum tensor of 11-dimensional
supergravity but instead with the tensor ${\cal T}^\mu{}_a$, see
eq.~({\ref{2ndduality}), involving the dual graviton and 6-form
together with their respective shift gauge fields, it is is not
possible to find a local expression for $Y^{a}$ that solves
(\ref{2ndduality}). To show this one may gauge-fix $C$ and $A_{(6)}$
away, after which this tensor symbolically reads ${\cal T} \ \sim\
Y^2+Z^2$, such that one is left with a $Z^2$ term that cannot be
written as the derivative of some local expression.

We conclude that, unlike the 3-form/6-form sector, the dual gravity
sector does not allow the elimination of the shift gauge field $Y$.
It is therefore impossible to generate the transformation rule
\eqref{dualsym} predicted by $E_{11}$ for the dual graviton as the
result of a compensating gauge transformation in the absence of $Y$.

\section{Conclusions}
In this letter we addressed the problem of reconciling the dual
graviton with $E_{11}$. For this we used the recently proposed
reformulation of gravitational theories \cite{Boulanger:2008nd},
which involves the dual graviton and circumvents the no-go results
of \cite{Bekaert:2002uh,Bergshoeff:2008vc} by virtue of keeping the
original graviton via a topological term in such a way that upon
linearization the spin-2 degrees of freedom can be encoded either in
the graviton or its dual. This reformulation contains all fields
required by $E_{11}$ (up to the given level) and allows to encode
the field equations in terms of an enlarged set of first-order
`duality relations'. Though property (i) discussed in the
introduction is therefore maintained, in contrast to the p-form
sector this is a priori independent of the validity of requirement
(ii) according to which this reformulation should realize the
$E_{11}$ symmetry on the dual fields. In answering the question
whether (a truncation of) $E_{11}$ is a symmetry of 11-dimensional
supergravity, a crucial role is played by the local shift symmetry
that is realized on the dual fields. As a consequence of this local
shift symmetry the dual fields capture the most general symmetries
in that any supposed gauge invariance of 11-dimensional supergravity
has to result from this shift symmetry by a gauge-fixing. The
question whether 11-dimensional supergravity does or does not
non-trivially realize the symmetry (\ref{dualsym}) on the dual
graviton predicted by $E_{11}$ can therefore be made precise by
asking whether there is a gauge-fixing that eliminates the shift
gauge field such that the residual gauge invariance gives rise to
the symmetry rule predicted by $E_{11}$. We find that this is not
possible. In this sense $E_{11}$ by itself is not a symmetry of
11-dimensional supergravity, but can at most be part of an extended
symmetry structure going beyond $E_{11}$ and comprising the
additional St\"uckelberg fields.

It is interesting to compare our results with a recent exploration
of the way in which the local symmetries of supergravity can be
obtained from the global symmetries of non-linear realisations of
$E_{11}$  \cite{Riccioni:2009hi}. In \cite{Riccioni:2009hi}  the
transition from global to local symmetries was implemented by
introducing on top of each $E_{11}$ generator in the Borel
subalgebra an infinite set of so-called Ogievetsky generators. In
the case of p-forms the fields associated to these Ogievetsky
generators parameterize higher derivatives of the p-form fields,
thus extending the global symmetry to a local one without
introducing new fields. This situation changes for the mixed
symmetry generators like the one corresponding to the dual graviton
considered in this letter. In that case not all Ogievetsky
generators can be eliminated. In particular, one is left with a
curvature for the dual graviton, linear in derivatives, that
contains a `shift' gauge field corresponding to one of the
Ogievetsky generators. This resonates with our approach. Our results
therefore suggest that all fields associated to the Ogievetsky
generators, except for the shift gauge field $Y$ occurring in the
dual graviton curvature $G$, can be expressed in terms of
derivatives of the basic dual graviton curvature. Even though it
seems to be difficult to reconcile the extended symmetry structure
of \cite{Riccioni:2009hi} with the original $E_{11}$ beyond the
Borel subalgebra, it might be worth to investigate it in light of
the present model and its possible relation to supergravity and/or
M-theoretic extensions.

\section*{Acknowledgments} For useful comments and discussions we
would like to thank N.~Boulanger and P.~Sundell. This work is part
of the research programme of the `Stichting voor Fundamenteel
Onderzoek der Materie (FOM)'.

\end{document}